# Electrical detection of the spin polarization due to charge flow in the surface state of the topological insulator $Bi_{1.5}Sb_{0.5}Te_{1.7}Se_{1.3}$

Yuichiro Ando[1,2*†], Takahiro Hamasaki[2*], Takayuki Kurokawa[2], Kouki Ichiba[2], Fan Yang[3], Mario Novak[3], Satoshi Sasaki[3], Kouji Segawa[3], Yoichi Ando[3§], and Masashi Shiraishi[1¶]

1. Department of Electronic Science and Engineering, Kyoto University, Kyoto 615-8531, Japan
2. Graduate School of Engineering Science, Osaka University, Toyonaka, Osaka 560-8531, Japan
3. Institute of Scientific and Industrial Research, Osaka University, Ibaraki, Osaka 567-0047, Japan

## Abstract

We detected the spin polarization due to charge flow in the spin non-degenerate surface state of a three dimensional topological insulator by means of an all-electrical method. The charge current in the bulk-insulating topological insulator $Bi_{1.5}Sb_{0.5}Te_{1.7}Se_{1.3}$ (BSTS) was injected/extracted through a ferromagnetic electrode made of $Ni_{80}Fe_{20}$, and an unusual current-direction-dependent magnetoresistance gives evidence for the appearance of spin polarization which leads to a spin-dependent resistance at the BSTS/$Ni_{80}Fe_{20}$ interface. In contrast, our control experiment on $Bi_2Se_3$ gave null result. These observations demonstrate the importance of the Fermi-level control for the electrical detection of the spin polarization in topological insulators.

Three-dimensional (3D) topological insulators (TIs) represent a new quantum state characterized by topologically-protected gapless surface states with massless Dirac fermions [1-4]. One of the most prominent properties of the TI surface state is spin-momentum locking, i.e., spin quantization axis of the conduction electron in the TI surface state is perpendicularly locked to the carrier momentum as schematically shown in Fig. 1(a) [4]. Due to this spin-momentum locking, dissipationless pure spin currents exist in the TI surface state in thermal equilibrium [4]. Furthermore, it is expected that charge current naturally induces spin polarization, whose axis and the sign can be controlled by the direction of the charge flow.

The spin texture in the Dirac-cone of TI surface state has been investigated by means of spin- and angle-resolved photoemission spectroscopy (ARPES), which confirmed the spin-momentum locking in the TI surface state [5-8]. In order to realize spintronic devices using 3D TIs based on their spin-momentum-locked characteristics, extractions of the spin polarized currents from the TI surface state and their detections by all-electrical methods are desirable [9-14]. Recently, a successful detection of the charge-current-induced spin polarization on the TI surface state based on the measurements of the conventional spin accumulation voltage has been reported by Li et al [11]. Intriguingly, their study was implemented by using $Bi_2Se_3$ (BS), whose bulk state is not really insulating [11]; therefore, they needed to use very thin films to enhance the contribution of the surface state to the total conductance. In fact, Li et al. reported [11] that the spin accumulation voltage diminishes quickly with increasing thickness in their BS devices, and the maximum thickness was 45 nm. Hence, utilization of more bulk-insulating TIs would allow more efficient spin detection. Also, it would be useful to detect the spin polarization with a method other than the spin accumulation voltage measurements.

In this Letter, we report successful detection of the spin polarization due to charge flow in a bulk-insulating TI, $Bi_{1.5}Sb_{0.5}Te_{1.7}Se_{1.3}$ (BSTS), by using magnetoresistance measurement whose principle is different from the spin accumulation voltage measurements [11]. The magnetotransport studies of BSTS have been used to demonstrate the surface-dominated transport in this compound [15, 16], and the ARPES study has confirmed its Fermi level to be located in a bulk band gap, realizing the intrinsic insulating state [17]. In the present magnetoresistance measurements, we observed a rectangular hysteresis behavior which is governed by the resistance at the interface between BSTS and the ferromagnetic $Ni_{80}Fe_{20}$ (Py) electrode used for current injection/extraction; our data indicate that at 4.2 K the interface resistance changes both with the magnetization direction of Py and with the current direction. This peculiar magnetoresistance disappears when BS is used in the devices instead of BSTS, or when the BSTS device is heated to 300 K.

### Fabrication of TI-based Spin Devices

Figure 1(b) shows a scanning electron microscope (SEM) image of a typical TI device. Single crystals of BSTS and BS were grown by a Bridgeman method in evacuated quartz tubes [15, 16]. Mechanically exfoliated TI flakes, with the thickness of several tens of nm, were put on a thermally-oxidized $SiO_2$ (500 nm in thickness) layer formed on a Si substrate. The actual thickness of the TI flakes were measured by a combination of laser microscope and atomic force microscope. For magnetoresistance measurements, two nonmagnetic Au/Cr electrodes and several Py electrodes were fabricated by using electron beam lithography and electron beam evaporation. The width of the Py electrodes was 500 – 800 nm. Charge currents are injected/extracted between a Au/Cr electrode and a Py electrode, which causes a spin accumulation at the TI/Py interface; it is because of the current-induced spin polarization in the TI surface due to the spin momentum locking. As a result, the interface resistance is expected to become dependent on the magnetization direction of the Py layer as well as the current direction. We also made standard Hall-bar devices with six nonmagnetic Au/Cr electrodes using the BSTS flakes on the same wafer, which were used for characterizing the resistivity of the TI channel in the magnetoresistance devices.

### Temperature dependence of resistivity of exfoliated TI channels

The temperature dependences of the resistivity of three of our TI samples, whose cross-sectional area $S = tw$ is given by the thickness $t$ and the width $w$ of the flakes, is shown in Fig. 1(c). The resistivity of the BSTS flakes, $\rho_{BSTS}$, used for this resistivity characterization are essentially reproducible and always presents weakly semiconducting behavior above ~50 K, although the behavior at lower temperature was sample dependent, presumably due to different levels of disorder. Nevertheless, thanks to the topological protection of the surface state [18], our BSTS samples always present plateau-like resistivity behavior at sufficiently low temperature [below ~75 K (50 K) in the $t$ = 45 nm (58 nm) sample shown in Fig. 1(c)]. The observed behavior is essentially the same as that in bulk BSTS crystals [15, 16], indicating that the electrical properties are maintained after exfoliation. In contrast, the resistivity of the BS flakes, $\rho_{BS}$, is much smaller than $\rho_{BSTS}$ and presents metallic behavior throughout the measured temperature range, which comes from the dominance of the degenerate bulk transport channel [12] due to the Fermi level $E_F$ located above the conduction band bottom [19].

**Electrical detection of the spin polarization in the surface state of $Bi_{1.5}Sb_{0.5}Te_{1.7}Se_{1.3}$.**

Magnetoresistance measurements were performed in a multi terminal scheme employing ferromagnetic contacts [9, 10] to detect the spin polarization on the TI surface. In this scheme, a DC current $I$ is applied between contacts 1 and 2, whereas the voltage is measured between contacts 2 and 3; since there is no current flow between contacts 2 and 3, this measurement reads only the voltage drop occurring at the contact 2, $V_2$, see Fig 1(b) [20-23]. When the contact 2 is made of a ferromagnet, a spin-dependent resistance can be detected through $V_2$, because the charge current in the TI surface state is spin polarized due to the spin-momentum locking. It is noted that magnetoresistance in this study is caused by spin polarized current injected into / extracted from TI surface state by means of an electric field, which is different from previous studies [11, 12]. (see Supporting Information S1) The interfacial resistance $R_2$ (= $V_2/I$) as a function of the in-plane magnetic field $H$ for a BSTS device (Device A) at $I$ = 100 μA, measured at 4.2 K, is shown in Fig. 2(a). A rectangular hysteresis feature with steep resistance changes at ±150 Oe can be recognized, as indicated by dotted arrows; here, $R_2$ becomes higher when the magnetization of the Py electrode is aligned along the -$x$ direction. Importantly, when the polarity of $I$ is reversed, the pattern of the rectangular hysteresis feature reverses, as shown in Fig. 2(b), where $R_2$ now becomes lower when the magnetization is aligned along the -$x$ direction.

Moreover, the reversal of the rectangular hysteresis feature is also observed when the current-voltage scheme is changed as shown in Figs. 2(c) and 2(d). Such a behavior indicates that the direction of the spin polarization reverses with the current direction, which is a natural consequence of the spin-momentum locking in the surface state. In this respect, it is important to note that the sign of the spin-dependent signal is consistent with the left-handed helicity of the topological surface state of BSTS above the Dirac point [4]. Furthermore, we have also measured $R_2$ vs $H$ for the case when $H$ is parallel to the current. As shown in Fig. S4 of the Supporting Information, the rectangular hysteresis is not observed in this geometry, which strongly support our conclusion.

In passing, since the dip features marked by thick arrows in Figs. 2(a) and 2(b) are attributed to the anisotropic magnetoresistance (AMR) of the Py electrode, one can be confident that the rectangular hysteresis feature is not due to the AMR effect of the Py electrode. This rectangular hysteresis behavior was reproduced in several BSTS devices. Therefore, it is reasonable to conclude that the charge-current-induced spin polarization is detected at the TI/ferromagnet interface in terms of a peculiar magnetoresistance feature.

**Temperature and Current dependences of the spin signals.**

The temperature dependence of the magnitude of the spin signal was also investigated in devices A and C. The $R_2$ vs. $H$ curve measured at 300 K in device A at $I$ = +100 μA is shown in Fig. 3(a). The rectangular hysteresis behavior has disappeared, whereas the AMR signal is kept being observed. In device C, neither the rectangular hysteresis nor the AMR signal was observed at 300 K [the inset of Fig. 3(b)]; in fact, the temperature dependence of the magnitude of the spin signals, $\Delta V_2$, in device C [main panel of Fig. 3(b)] indicates that $\Delta V_2$ decreases monotonically with increasing temperature and vanishes at around 150 K. Although the bulk conduction usually becomes dominant at 300 K even in a bulk-insulating TI, in our devices the fraction of the surface conductance in the total conductance is expected to be not reduced significantly, because the total resistance of our BSTS flakes changes by less than a factor of 1.3 between 4.2 and 300 K. Also, it is useful to note that, whereas the entering into the surface-dominated transport regime occurs at 50–100 K in BSTS [15], $\Delta V_2$ keeps increasing with decreasing temperature even below 50 K [see Fig. 3(b)]. Therefore, the disappearance of the spin signal in BSTS should be due to a mechanism other than the dominance of the bulk transport. In this regard, a possible origin is the reduction in the spin polarization of Py [24-26], but it would be useful to test other ferromagnets or to insert a tunneling barrier for elucidating the key factor to prohibit the detection of the spin polarization at room temperature in the present devices. In passing, the difference in the observability of the AMR signal between samples A and C is most likely due to a difference in the shape of the Py electrode; remember, when a Py electrode is sharp and smooth, the shape magnetic anisotropy becomes stronger, which leads to a steep magnetization change. As a result, the AMR signal whose origin is resistance reduction when the magnetization direction is perpendicular to the charge-current direction is diminished.

The dependence of $\Delta V_2$ on current $I$ in the device B is shown in Fig. 3(c). One can see that the relationship is essentially linear. From the spin-charge coupled transport equations given by Burkov and Hawthorn [9], magnetoresistance in the FM/ TI surface state /NM structure is expressed as

$$\Delta V_2 = \frac{8\pi\hbar\eta I}{e^2 k_F w} \quad , \qquad (1)$$

where $\eta$ is the spin polarization of the injected/extracted current from the Py electrode, $\hbar$ is the reduced Planck constant, $e$ is the elementary charge, $k_F$ is the Fermi wavenumber, and $w$ is the width of the TI channel. Since $\Delta V_2$ is proportional to $I$ in Eq. (1), our result in Fig. 3(c) is consistent with the theoretical expectation. When we use the value $k_F \approx 0.1$ Å$^{-1}$ for BSTS [27], $\eta$ is calculated to be 0.05~0.5 %. Such a small $\eta$ is reasonably understood as a result of the conductance mismatch problem between the metallic Py electrode and the relatively-high-resistance TI channel, which leads to a small magnetoresistance [28]. Note that a diffusive transport is assumed in the model used here, and the current-induced spin density in the TI surface state is given by $I/(2ev_F)$ [9].

As a control experiment, we also measured the magnetoresistance of many BS-based devices. The $R_2$ vs. $H$ curves of a BS device measured at 4.2 and 300 K are shown in Figs. 4(a)–4(d) for both current polarities. One can see that no rectangular hysteresis feature was observed irrespective of the temperature, although AMR signals were observed at both low and high temperatures. Whilst the magnetoresistance measurements were performed for more than ten BS devices, no rectangular hysteresis signals were observed, which is due to the dominant bulk conduction at all temperatures. Clearly, to observe a sizable spin-related signal in BS devices, the thickness should be less than 30 nm [11], which is difficult to achieve with exfoliated flakes.

In summary, detection of the charge-current-induced spin polarization has been demonstrated by using the spin-dependent magnetoresistance at a TI/Py interface, for which the bulk-insulating TI $Bi_{1.5}Sb_{0.5}Te_{1.7}Se_{1.3}$ was used for enhancing the fraction of the surface current. Control experiments using $Bi_2Se_3$-based devices have clarified the importance of the Fermi-level control for the electrical detection of the spin polarization in topological insulators.

**Author Information**

*Equal contribution.

† ando@kuee.kyoto-u.ac.jp

§ y_ando@sanken.osaka-u.ac.jp

¶ mshiraishi@kuee.kyoto-u.ac.jp

The authors declare no competing financial interest.

**Acknowledgements**

This work was supported by JSPS (KAKENHI No. 25220708), MEXT (Innovative Area "Topological Quantum Phenomena" KAKENHI Nos. 22103004 and 25103715), and AFOSR (AOARD 124038).

## Figure captions

**Figure 1**

**Device structure of TI-based spin devices and electrical properties of TI.**

(a) Schematic illustrations of the spin-momentum-locked surface state of 3D TIs: energy dispersion with the spin quantization axes shown by arrows (left); real-space image of the momentum direction and the spin quantization axes (right); relationship between charge and spin currents (bottom). (b) A false-color SEM image of a typical TI device for the spin-dependent magnetoresistance measurements. (c) Temperature dependences of the resistivity of exfoliated BSTS and BS flakes.

**Figure 2**

**Electrical detection of the spin polarization in the surface state of $Bi_{1.5}Sb_{0.5}Te_{1.7}Se_{1.3}$.**

(a, b) Magnetic-field (*H*) dependences of the interface resistance $R_2$, measured at 4.2 K in BSTS device A at bias currents of (a) +100 μA and (b) -100 μA; the thickness of the BSTS flake was 34 nm. (c, d) $R_2$ vs *H* curves for device B made with a 23 nm-thick BSTS flake, measured at 4.2 K for $I$ = -100 μA. The charge current was applied from contact 2 to contact 1 in (c), and from contact 2 to contact 3 in (d).

**Figure 3**

**Temperature and current dependence of the spin signal.**

(a) $R_2$ vs *H* curves of device A at $I$=+100 μA measured at 300 K. (b) Temperature dependence of $\Delta V_2$ in device C (67 nm thick) at $I$=+ 50 μA. The upper and lower insets show $R_2$ vs *H* curves measured at 4.2 and 300 K, respectively. (c) $I$ dependence of the magnitude of the spin signal, $\Delta V_2$, of device B measured at 4.2 K.

**Figure 4**

**Magnetoresistance in $Bi_2Se_3$-based device**

$R_2$ vs *H* curves of a 72 nm-thick BS device measured at 4.2 K for (a) $I$ = +100 μA and (b) $I$ = -100 μA, and at 300 K for (c) $I$ = +100 μA and (d) $I$ = -100 μA, respectively.

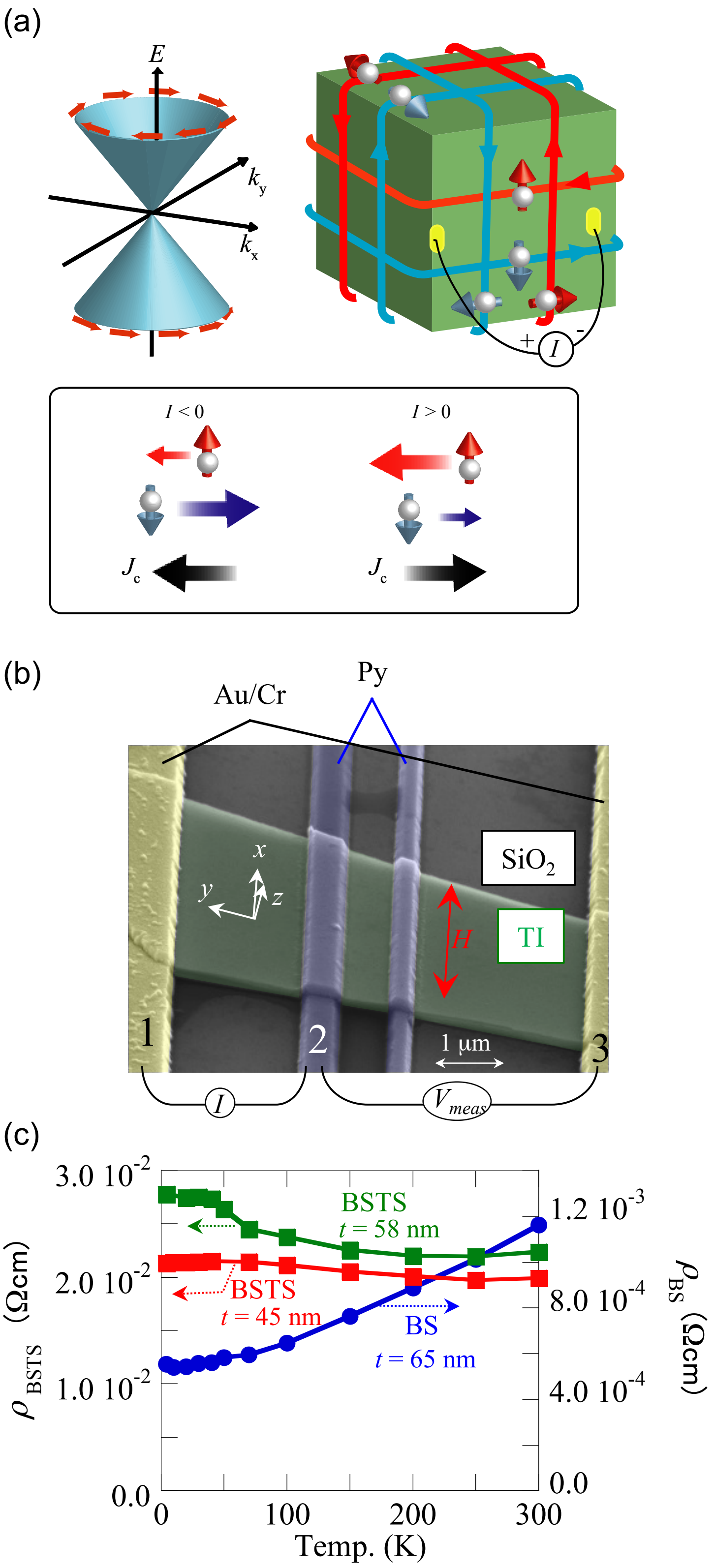

Fig 1 Ando Hamasaki

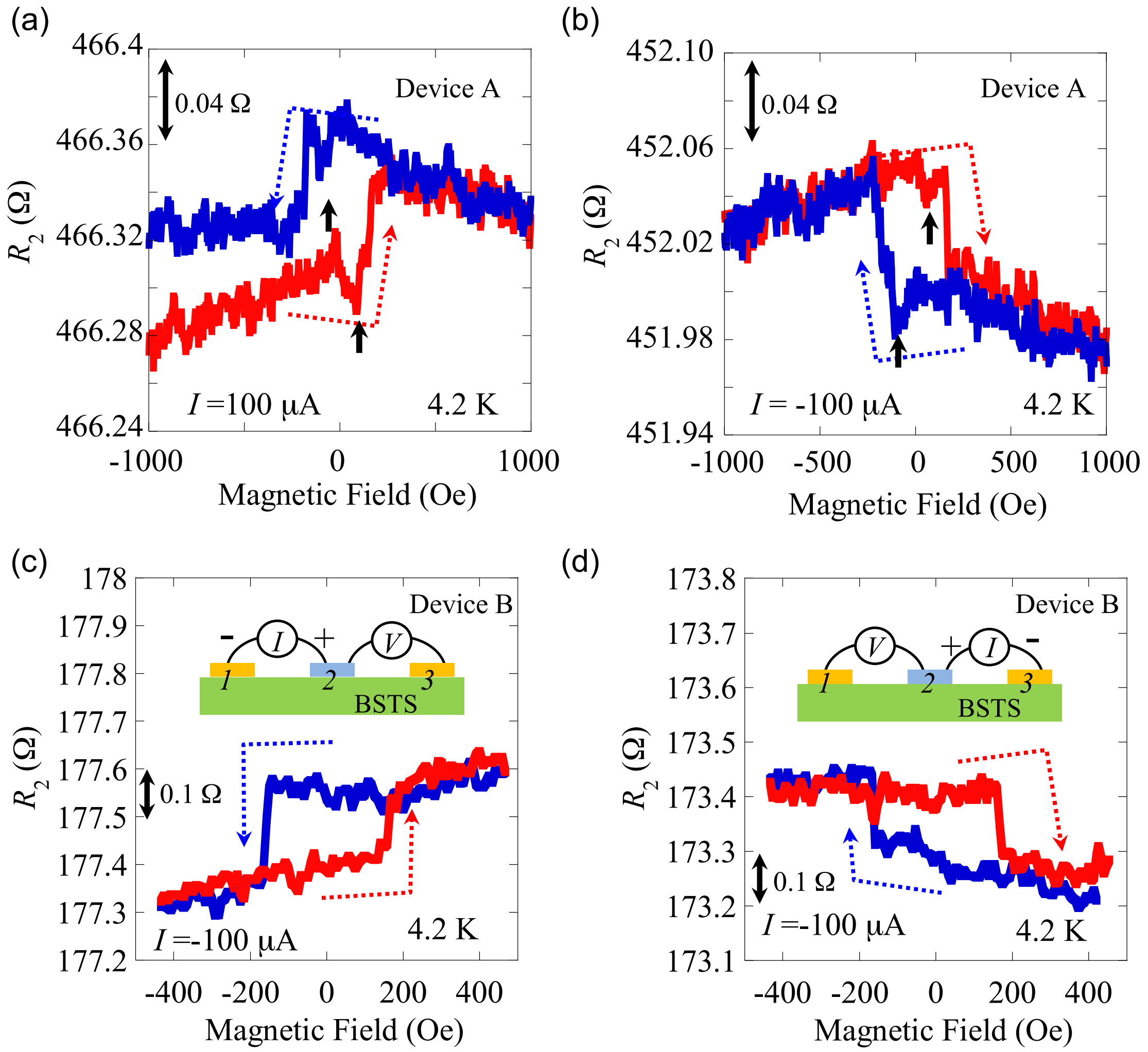

Fig 2 Ando Hamasaki

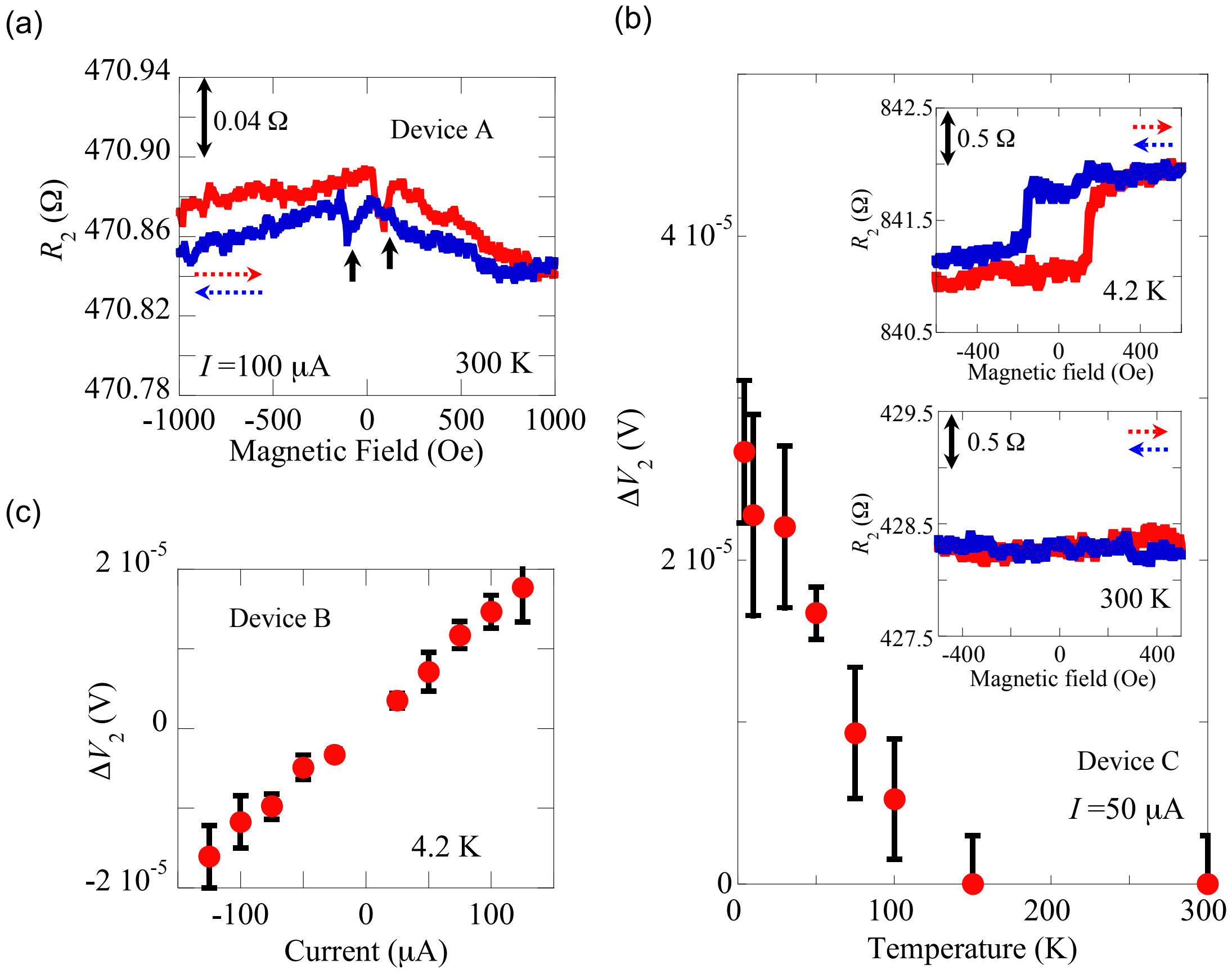

Fig 3 Ando Hamasaki

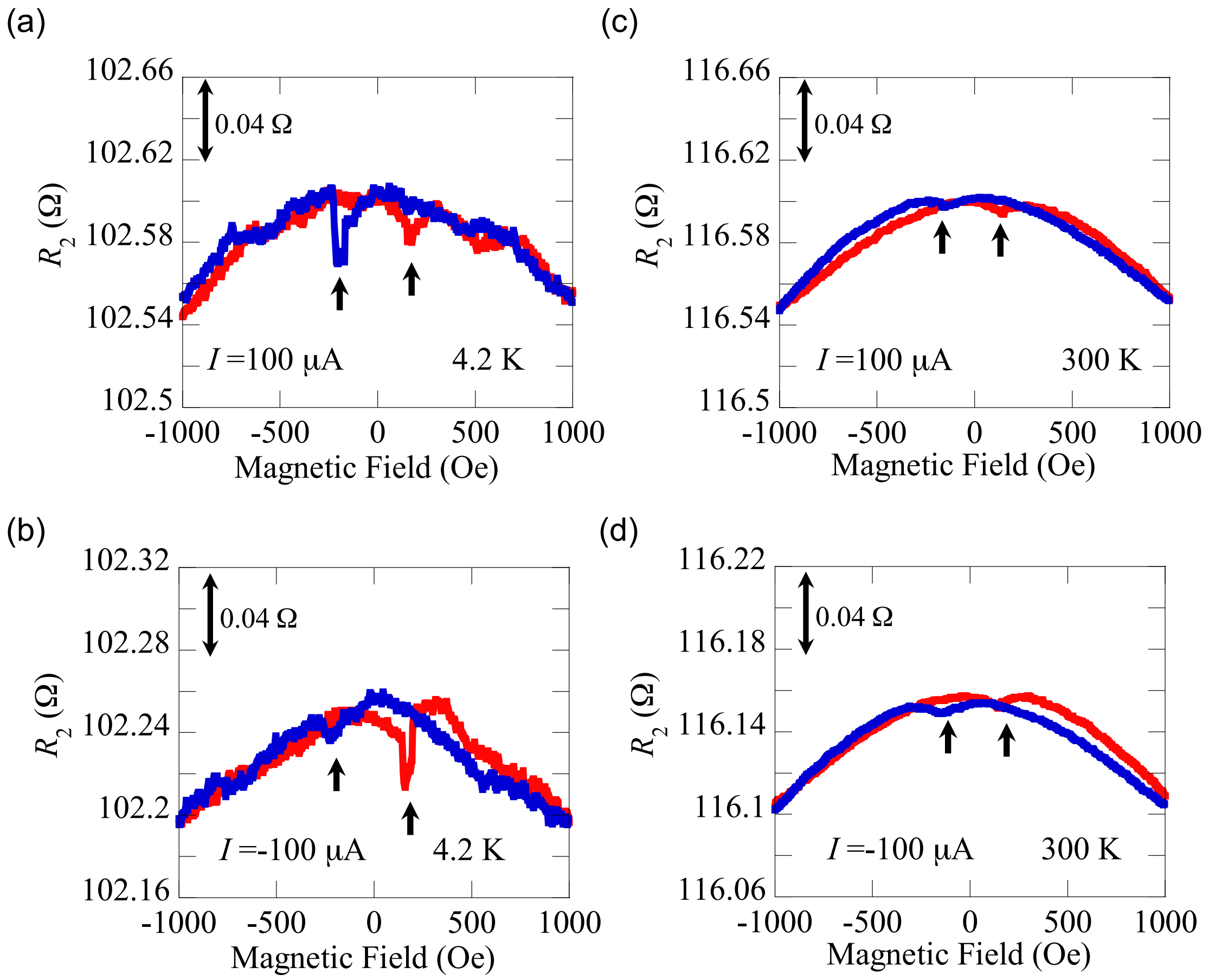

Fig. 4 Ando Hamasaki